\documentclass[prb,onecolumn,aps,showpacs]{revtex4}
\usepackage{graphicx}%
\usepackage{dcolumn}
\usepackage{amsmath}
\usepackage{amssymb}
\usepackage{epsfig}

\begin{document}

\title{Speed of sound in a superfluid Fermi gas in an optical
lattice}
\author{Z. G. Koinov}\affiliation{Department of Physics and Astronomy,
University of Texas at San Antonio, San Antonio, TX 78249, USA}
\email{Zlatko.Koinov@utsa.edu} 
\begin{abstract}
A system of equal mixture of $^6Li$ atomic Fermi gas of two
hyperfine states loaded into a cubic three-dimensional optical
lattice is studied assuming a negative scattering length (BCS side
of the Feshbach resonance). When the interaction is attractive,
fermionic atoms can pair and form a superfluid. The dispersion of
the phonon-like mode and the speed of sound in the long-wavelength
limit are obtained by solving the Bethe-Salpeter equations for the
collective modes of the attractive Hubbard Hamiltonian.
\end{abstract}
\maketitle

\section{Introduction}
In the last decade the possibility of a superfluid alkali atom Fermi
gas has attracted much attention  both theoretically
\cite{SF1,SF2,SF3,SF4,SF5,SF6,SF7,SF8,SF9,SF10,SF11,SF12,SF13,SF14}
and experimentally \cite{SFexp} because this phenomenon opens a new
opportunity to study strongly correlated quantum many-particle
systems and to emulate high-temperature superconductors. Optical
lattices are made with lasers, and therefore, the lattice geometry
is easy to modify by changing the wavelength of the intersecting
laser beams.  Near the Feshbach resonance the atom-atom interaction
can be manipulated in a controllable way because the scattering
length $a_s$ can be changed from the BCS side (negative values) to
the BEC side (positive values) reaching very large values close to
resonance. We focus our attention on the BCS transition (negative
scattering length) of degenerate fermionic gases to a superfluid
state analogous to superconductivity. In particular, we consider an
equal mixture of $^6Li$ atomic Fermi gas of two hyperfine states
$|F=1/2, m_f=\pm 1/2>$ with contact interaction loaded into an
optical lattice. The two hyperfine states are described by
pseudospins $\sigma=\uparrow,\downarrow$. We also assume that the
number of atoms in each hyperfine state per site (the filling
factor) is smaller than unity, and that the lattice potential is
sufficiently deep such that the tight-binding approximation is
valid. The system in this case is well described by the single-band
Hubbard model:
\begin{equation}H=-J\sum_{<i,j>,\sigma}\psi^\dag_{i,\sigma}\psi_{j,\sigma}
-\mu\sum_{i,\sigma}\widehat{n}_{i,\sigma}+U\sum_i
\widehat{n}_{i,\uparrow} \widehat{n}_{i,\downarrow}.
\label{Hubb1}\end{equation} Here, the Fermi operator
$\psi^\dag_{i,\sigma}$ ($\psi_{i,\sigma}$) creates (destroys) a
fermion on the lattice site $i$ with pseudospin
$\sigma=\uparrow,\downarrow$ and
$\widehat{n}_{i,\sigma}=\psi^\dag_{i,\sigma}\psi_{i,\sigma}$ is the
density operator on site $i$ with a position vector $\textbf{r}_i$.
$\mu$ is the chemical potential, and the symbol $\sum_{<ij>}$ means
sum over nearest-neighbor sites. $J$ is the tunneling strength of
the atoms between nearest-neighbor sites, and $U$ is the on-site
interaction. On the BCS side the interaction parameter $U$ is
negative (the atomic interaction is attractive). For simplicity we
assume that each well of the periodic potential for atomic motion in
three dimensions could be approximated by a harmonic potential. This
harmonic approximation gives the following analytical results of $J$
and $U$ ($\hbar=1$) \cite{SF6}:
$$J= E_R e^{-\frac{\pi^2\sqrt{s}}{4}}\left[\frac{\pi^2s}{4}-\frac{\sqrt{s}}{2}
-\frac{s}{2}\left(1+e^{-\sqrt{s}}\right)\right],\quad
 U=-\frac{8}{\sqrt{\pi}}\frac{|a_s|}{\lambda} \left(\frac{2s^3E^3_R}{m\lambda^2}\right)^{1/4}.$$
Here $\lambda$ is the laser wavelength, $s$ is the lattice height,
and $m$ is the mass of the trapped $^6Li$ atoms. The recoil energy
of the lattice $E_R=\pi^2/2md^2$ depends on the lattice constant
$d=\lambda/2$. In our numerical calculations the wavelength  is
chosen to be $\lambda=1030$ nm ($E_R=1.293\times 10^{-11}$ eV)
\cite{SF6}.  The lattice height is assumed to be $s=2.5$.

In what follows we study the spectrum of the collective modes of the
Hamiltonian (\ref{Hubb1}). According to the Goldstone theorem, the
long-wavelength limit of the spectrum has to be linear which means
the speed of sound in this limit is independent of the wave-vector.
In Ref. [\onlinecite{SF6}] the spectrum of the collective modes has
been obtained from the poles of the density response function which
 had been calculated in the generalized random phase approximation
 (GRPA). This response-function version of the GRPA uses a
 $4\times 4$ matrix $L^0$ (we follow the
 notations used in Ref. [\onlinecite{SF6}]) which has nine (not six as it
 is stated in Ref. [\onlinecite{SF6}]) independent elements:
 $a$, $b$, $c$, $\overline{c}$, $d$, $\overline{d}$, $L_{1222}$,
 $L_{2212}$ and $L_{2222}$. Thus, the response-function version of the GRPA has produced
 incorrect expressions for the density response function (see Eqs.
 (26) and (27) in Ref. [\onlinecite{SF6}]). At zero temperature, the correct GRPA leads
 to the following Bethe-Salpeter (BS) equations for the collective mode
$\omega(\textbf{Q})$ and corresponding BS amplitudes
$G^{\pm}(\textbf{k},\textbf{Q})$ \cite{ZK}:
\begin{equation}\begin{split}&[\omega(\textbf{Q})-\varepsilon(\textbf{k},\textbf{Q})]
G^{+}(\textbf{k},\textbf{Q})=\\& \frac{U}{2N}\sum_{\textbf{q}}
\left[\gamma_{\textbf{k},\textbf{Q}}\gamma_{\textbf{q},\textbf{Q}}+
l_{\textbf{k},\textbf{Q}}l_{\textbf{q},\textbf{Q}}\right]
G^{+}(\textbf{q},\textbf{Q})-\frac{U}{2N}\sum_{\textbf{q}}
\left[\gamma_{\textbf{k},\textbf{Q}}\gamma_{\textbf{q},\textbf{Q}}-
l_{\textbf{k},\textbf{Q}}l_{\textbf{q},\textbf{Q}}\right]
G^{-}(\textbf{q},\textbf{Q})\\&-\frac{U}{2N}\sum_{\textbf{q}}
\widetilde{\gamma}_{\textbf{k},\textbf{Q}}\widetilde{\gamma}_{\textbf{q},\textbf{Q}}
\left(G^{+}(\textbf{q},\textbf{Q})
-G^{-}(\textbf{q},\textbf{Q})\right)+\frac{U}{2N}\sum_{\textbf{q}}m_{\textbf{k},\textbf{Q}}m_{\textbf{q},\textbf{Q}}
\left[G^{+}(\textbf{q},\textbf{Q})+G^{-}(\textbf{q},\textbf{Q})\right],
\label{NewEq1} \end{split}\end{equation}
\begin{equation}\begin{split}
&[\omega(\textbf{Q})+\varepsilon(\textbf{k},\textbf{Q})]G^{-}(\textbf{k},\textbf{Q})=\\&
-\frac{U}{2N}\sum_{\textbf{q}}
\left[\gamma_{\textbf{k},\textbf{Q}}\gamma_{\textbf{q},\textbf{Q}}+
l_{\textbf{k},\textbf{Q}}l_{\textbf{q},\textbf{Q}}\right]
G^{-}(\textbf{q},\textbf{Q})+\frac{U}{2N}\sum_{\textbf{q}}
\left[\gamma_{\textbf{k},\textbf{Q}}\gamma_{\textbf{q},\textbf{Q}}-
l_{\textbf{k},\textbf{Q}}l_{\textbf{q},\textbf{Q}}\right]
G^{+}(\textbf{q},\textbf{Q})\\& -\frac{U}{2N}\sum_{\textbf{q}}
\widetilde{\gamma}_{\textbf{k},\textbf{Q}}\widetilde{\gamma}_{\textbf{q},\textbf{Q}}
\left(G^{+}(\textbf{q},\textbf{Q})
-G^{-}(\textbf{q},\textbf{Q})\right)-\frac{U}{2N}\sum_{\textbf{q}}m_{\textbf{k},\textbf{Q}}m_{\textbf{q},\textbf{Q}}
\left[G^{+}(\textbf{q},\textbf{Q})+G^{-}(\textbf{q},\textbf{Q})\right].
\label{NewEq2}
\end{split}\end{equation}
 Here the form factors are defined as follows:
$\gamma_{\textbf{k},\textbf{Q}}=u_{\textbf{k}}u_{\textbf{k}+\textbf{Q}}+v_{\textbf{k}}v_{\textbf{k}+\textbf{Q}},\quad
l_{\textbf{k},\textbf{Q}}=u_{\textbf{k}}u_{\textbf{k}+\textbf{Q}}-v_{\textbf{k}}v_{\textbf{k}+\textbf{Q}},
\quad
\widetilde{\gamma}_{\textbf{k},\textbf{Q}}=u_{\textbf{k}}v_{\textbf{k}+\textbf{Q}}-u_{\textbf{k}+\textbf{Q}}v_{\textbf{k}},
$ and $ m_{\textbf{k},\textbf{Q}}=
u_{\textbf{k}}v_{\textbf{k}+\textbf{Q}}+u_{\textbf{k}+\textbf{Q}}v_{\textbf{k}}
$ where $u^2_{\textbf{k}}=1-v^2_{\textbf{k}}=
\left[1+\overline{\varepsilon}(\textbf{k})/E(\textbf{k})\right]/2$.
The quantity $\varepsilon(\textbf{k},\textbf{Q})=
E(\textbf{k}+\textbf{Q})+E(\textbf{k})$, where
 $E(\textbf{k})=\sqrt{\overline{\varepsilon}^2_\textbf{k}+\Delta^2}$ depends
 on the gap
function  $\Delta$ and
 the mean-field electron energy $\overline{\varepsilon}_\textbf{k}$.
 We use a tight-binding form of the mean-field electron energy:
$\overline{\varepsilon}_\textbf{k}=2J\left(\cos k_xd+\cos k_yd+\cos
k_zd\right) -\mu$, where  $\mu$ is the chemical potential. The gap
function  and the chemical potential have to be determined by the
BCS number and gap equations:
\begin{equation}
1-f=\frac{1}{N}\sum_{\textbf{k}}\frac{\overline{\varepsilon}_\textbf{k}}{E(\textbf{k})},\quad
1=\frac{U}{N}\sum_{\textbf{k}}\frac{1}{2E(\textbf{k})},
\label{NGEq}\end{equation} where $f=M/N$ is the filling factor, and
we have $M$ atoms distributed along $N$ sites.

The BS equations for the collective modes can be reduced to a set of
four coupled linear homogeneous equations. The existence of a
non-trivial solution requires that the secular determinant
$det\|\widehat{\chi}^{-1}-\widehat{V}\|$ is equal to zero, where the
bare mean-field-quasiparticle response function $\widehat{\chi}$ and
the interaction $\widehat{V}=diag(-U,-U,U,-U)$ are $4\times 4$
matrices:\begin{equation} \widehat{\chi}=\left|
\begin{array}{cccc}
I_{\gamma,\gamma}&J_{\gamma,l}&I_{\gamma,\widetilde{\gamma}}&J_{\gamma,m}\\
J_{\gamma,l}&I_{l,l}&J_{l,\widetilde{\gamma}}&I_{l,m}\\
I_{\gamma,\widetilde{\gamma}}&J_{l,\widetilde{\gamma}}&
I_{\widetilde{\gamma},\widetilde{\gamma}}&
J_{\widetilde{\gamma},m}\\
J_{\gamma,m}&I_{l,m}&J_{\widetilde{\gamma},m}&I_{m,m}\end{array}%
\right|.\label{D}
\end{equation} Here we have introduced symbols
$I_{a,b}=F_{a,b}(\varepsilon (\mathbf{k},\mathbf{Q}))$ and
$J_{a,b}=F_{a,b}(\omega)$, where $F_{a,b}(x)$ is defined as follows (the quantities $a(\mathbf{k}%
,\mathbf{Q})$ and $b(\mathbf{k},\mathbf{Q})=l_{\mathbf{k},\mathbf{Q}},m_{%
\mathbf{k},\mathbf{Q}},\gamma _{\mathbf{k},\mathbf{Q}}$ or $\widetilde{%
\gamma }_{\mathbf{k},\mathbf{Q}}$):
$$
F_{a,b}(x)\equiv \frac{1}{N}\sum_\textbf{k}\frac{%
xa(\mathbf{k},\mathbf{Q})b(\mathbf{k},%
\mathbf{Q})}{\omega ^{2}-\varepsilon ^{2}(\mathbf{k},\mathbf{Q})}.$$

It is worth mentioning that the GRPA equations for the collective
mode derived by Belkhir and Randeria \cite{BR} can be obtained if we
neglect in (\ref{D}) all elements with index $\widetilde{\gamma}$.
In this case $\widehat{\chi}$ and $\widehat{V}$ are $3\times 3$
matrices.
\section{Speed of sound in a cubic lattice}
The velocity of sound is important because it tells us how fast the
sound propagates in the system, but more importantly, it is
intimately related to the normal (phonon) part of the liquid
according to Landau's theory of superfluidity \cite{L}.

\begin{figure}[tbp]
\includegraphics{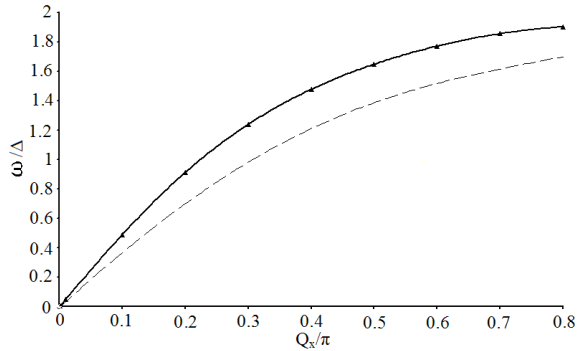} \label{Fig. 2}
\caption{The dispersion $\omega/\Delta$ of the phonon-like
collective mode. For filling factor $f=0.5$, lattice height $s=2.5$,
and scattering length $a_s=-1000 a_B$ ($a_B$ is the Bohr radius of
hydrogen), the chemical potential $\mu=0.326E_R$ and the gap energy
$\Delta=0.05E_R$ are obtained by solving the number and gap
equations (\ref{NGEq}). The speed of sound in the long-wavelength
limit is $8.1$ mm/s. The puncture curve represents the dispersion
calculated in Ref. [\onlinecite{SF6}].}\end{figure}
\begin{figure}[tbp]
\includegraphics{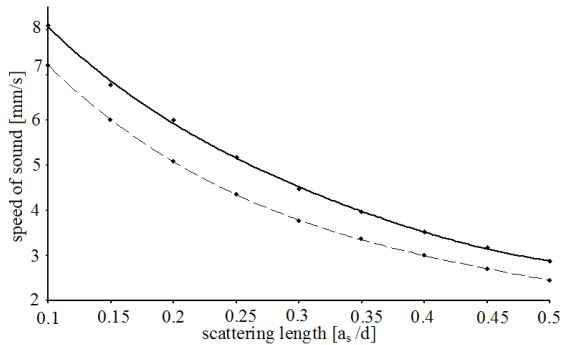} \label{Fig. 1}
\caption{The speed of sound (in units mm/s) as a function of the
scattering length $a_s/d$ ($d=515$ nm  is the lattice constant). The
filling factor is f=0.5, and the lattice height is $s=2.5$. The
puncture curve represents the speed of sound  calculated in Ref.
[\onlinecite{SF6}].}\end{figure}

 In our numerical calculations, the sum over
$\textbf{k}$ is replaced by a triple integral over the first
Brillouin zone: $-\pi\leq k_xd \leq\pi$, $-\pi\leq k_yd \leq\pi$ and
$-\pi\leq k_zd \leq\pi$. After that, we applied the substitutions
$x=\tan k_xd/4$,$y=\tan k_yd/4$ and $z=\tan k_zd/4$ to rewrite the
integrals in the form of Gaussian quadrature
$\int_{-1}^{1}dx\int_{-1}^{1}dy f(x,y,z)/(1+x)(1+y)(1+z)$. The
corresponding integrals are numerically evaluated using $49\times
49\times 49$ $(x_i,y_j,z_k)$ points:
$\int_{-1}^{1}dx\int_{-1}^{1}dy\int_{-1}^{1}dz
f(x,y,z)/(1+x)(1+y)(1+z)=\sum_{i=1}^{49}\sum_{j=1}^{49}\sum_{k=1}^{49}w_iw_jw_kf(x_i,y_j,z_k)$,
where $w_i$ is the corresponding weight. It can be checked that
there is no difference between the  approximation by integrals and
the case when the sums over $\textbf{k}$ are taken explicitly
assuming 128 sites per dimension.

In Fig. 1 and Fig. 2 we present the results of our calculations of
the dispersion of the phonon-like mode and the speed of sound as a
function of the scattering length assuming that the filling factor
and  the lattice height are $f=0.5$  and $s=2.5$, respectively. The
long-wavelength part of the dispersion is linear with sound velocity
of about $8.1$ mm/s. For higher momenta the dispersion saturates to
$2\Delta$. As it is expected, when the interaction between the atoms
is increased by increasing the scattering length, the
compressibility of the system increases, and therefore, the speed of
sound decreases, as can be seen in Fig. 2. In both figures, there
exists a difference of about 10 -15 percents between the BS approach
and the response-function calculations presented in Ref.
[\onlinecite{SF6}].
\section{Conclusion}
In this paper, we have used the BS equations in the GRPA to obtain
the dispersion of the phonon-like collective mode and the
corresponding sound velocity in the long-wavelength limit in the
system of equal mixture of $^6Li$ atomic Fermi gas of two hyperfine
states loaded into a cubic three-dimensional optical lattice. It is
shown that the previous calculations, which have been obtained by
studding the poles of the density response functions, are not in
accordance with our results derived by means of the BS equations in
the GRPA.

\end{document}